\documentclass[useAMS,usenatbib,twocolumn]{mnras}

\usepackage[]{color,graphicx}

\usepackage{amsmath}

\def\lesssim{\mathrel{\hbox{\rlap{\hbox{\lower5pt\hbox{$\sim$}}}\hbox{$<$}}}}
\def\gtrsim{\mathrel{\hbox{\rlap{\hbox{\lower5pt\hbox{$\sim$}}}\hbox{$>$}}}}

\title[$^{56}$Ni production of CCSN]
{Importance of $^{56}$Ni production on diagnosing explosion mechanism of core-collapse supernova}
\author[Suwa, Tominaga, and Maeda]{
Yudai Suwa$^{1}$\thanks{E-mail: suwa@yukawa.kyoto-u.ac.jp},
Nozomu Tominaga$^{2,3}$,
and 
Keiichi Maeda$^{4,3}$
\\
$^{1}$Center for Gravitational Physics, Yukawa Institute for Theoretical Physics, Kyoto University, Kyoto, 606-8502, Japan\\
$^{2}$Department of Physics, Faculty of Science, and Engineering, Konan University, 8-9-1 Okamoto, Kobe, Hyogo 658-8501, Japan\\
$^{3}$Kavli Institute for the Physics, and Mathematics of the Universe (WPI), University of Tokyo, Kashiwa, Chiba 277-8583, Japan\\
$^{4}$Department of Astronomy, Kyoto University, Oiwake-cho, Kitashirakawa, Sakyo-ku, Kyoto 606-8502, Japan}

\begin{document}

\date{Accepted. Received.}

\pagerange{\pageref{firstpage}--\pageref{lastpage}} \pubyear{2017}

\maketitle

\label{firstpage}

\begin{abstract}
$^{56}$Ni is an important indicator of the supernova explosions, which
  characterizes light curves. Nevertheless, rather than $^{56}$Ni, the
  explosion energy has often been paid attention from the explosion
  mechanism community, since it is easier to estimate from numerical
  data than the amount of $^{56}$Ni. The final explosion energy,
  however, is difficult to estimate by detailed numerical simulations
  because current simulations cannot reach typical timescale of
  saturation of explosion energy. Instead, the amount of $^{56}$Ni
  converges within a short timescale so that it would be a better
  probe of the explosion mechanism. We investigated the amount of
  $^{56}$Ni synthesized by explosive nucleosynthesis in supernova
  ejecta by means of numerical simulations and an analytic model. For
  numerical simulations, we employ Lagrangian hydrodynamics code in
  which neutrino heating and cooling terms are taken into account by
  light-bulb approximation. Initial conditions are taken from
  \cite{woos07}, which have 12, 15, 20, and 25 $M_\odot$ in zero age
  main sequence. We additionally develop an analytic model, which
  gives a reasonable estimate of the amount of $^{56}$Ni. We found
  that, in order to produce enough amount of $^{56}$Ni,
  $\mathcal{O}(1)$ Bethe s$^{-1}$ of growth rate of the explosion
  energy is needed, which is much larger than that found in recent
  exploding simulations, typically $\mathcal{O}(0.1)$ Bethe s$^{-1}$.
\end{abstract}

\section{Introduction}
\label{sec:intro}

The important product of supernova nucleosynthesis is $^{56}$Ni, which
drives supernova brightness. A typical amount of $^{56}$Ni by
canonical supernovae is estimated as $\mathcal{O}(0.01)M_\odot$
\citep{hamu03b,smar09},\footnote{A typical amount of $^{56}$Ni of
  nearby supernovae (1987A, 1993J, and 1994I) is $\approx $0.07
  $M_\odot$ \citep[e.g.,][]{arne89,woos94,iwam94}.} which can be
measured by exponential tail from the late light curve with low
ambiguity. In contrast, the explosion energy, which has been used as
an indicator of the explosion simulations, needs two observables
(light curve and spectrum) to be estimated, since it is interfered by
a product of ejecta mass and velocity.\footnote{More precisely, from
  light curve we can estimate the geometrical mean of diffusion
  timescale of photons and hydrodynamical timescale,
  $\sqrt{t_\mathrm{diff}t_\mathrm{hyd}}\sim\sqrt{M_\mathrm{ej}\kappa/v_\mathrm{ej}}$,
  where $M_\mathrm{ej}$ is the ejecta mass, $\kappa$ is opacity, and
  $v_\mathrm{ej}$ is the typical velocity of the ejecta
  \citep{arne82}. The velocity can be independently measured by the
  spectrum. By assuming the opacity with a reasonable value
  $\kappa\approx 0.1$ cm$^{2}$ g$^{-1}$, we can resolve the degeneracy
  between mass and velocity.} It implies that the amount of $^{56}$Ni
has a smaller systematic error compared to the explosion
energy. Indeed, for SN 1998bw as an example, the estimated explosion
energy ranges from 2 to 25 Bethe (1 Bethe $\equiv 10^{51}$ erg)
\citep{hoef99,naka01c,maed06}, depending on details of radiation
transfer simulations and the ejecta structure assumed in such models,
and methods to derive the physical quantities from observables. On the
other hand, the estimated amount of $^{56}$Ni is converged between 0.2
and 0.4 $M_\odot$. In addition, production of $^{56}$Ni has been
suggested to be sensitive to the explosion mechanism, that is, the
energy deposition rate rather than the total explosion energy itself
\citep[see, e.g.][]{maed09,suwa15a}.

The mechanism of supernova explosions is still under a thick veil,
even though it has been already more than 80 years from the original
idea by \cite{baad34}, more than 50 years from the first numerical
simulation \citep{colg66}, and more than 30 years from the first
simulation of delayed explosion \citep{beth85}, which is the current
standard scenario of supernova explosion mechanism.

After a few decades of unsuccessful explosion era \citep{ramp00,
  lieb01, thom03, sumi05}, we have some exploding simulations since
\cite{bura06} \citep[e.g.][]{mare09, suwa10, taki12, muel12b, brue13,
  naka15, lent15, muel15b, pan16, ocon15,burr16}, in which
multidimensional hydrodynamics equations are solved simultaneously
with spectral neutrino transport. However, most of simulations have
been performed in two-dimension (with axial
symmetry). Three-dimensional simulations without any spacial symmetry
employed have shown worse results than two dimensional ones
\citep{hank12, couc13b, taki14, lent15}, since three dimensional
turbulence leads to an energy cascade from large scale to small scale
(normal cascade), while two dimensional one makes it opposite (inverse
cascade). It is known that a large scale, i.e. global, turbulence aids
the explosion, so that some results from two-dimensional simulations
might reflect a numerical artifact and these simulations might well
overestimate the explosion energy.

The state-of-the-art simulations have shown slow increase of the
explosion energy. As summarized in Table \ref{tab:models}, the growing
rate of the explosion energy is typically $\mathcal{O}(0.1)$ Bethe
s$^{-1}$, especially for 3D simulations.  Therefore, it can be argued
that, by neutrino heating mechanism, these simulations require at
least a few second to get a canonical explosion energy, i.e. 1
Behte.\footnote{These simulations are all starting from stellar
  evolutionary results. By changing initial condition, the growth rate
  of the explosion energy can be $\approx 5$ Bethe s$^{-1}$ even in
  spherical symmetry \citep{suwa16b}.} It should be noted that the
explosion energy estimated in explosion simulations is not a direct
observable, since there is bound (totally negative energy) material
above the shock and it reduces the explosion energy when it is swept
by the shock.

The explosion energy is related to the $^{56}$Ni synthesis, since to
synthesize $^{56}$Ni the temperature needs to be $T\gtrsim 5\times
10^9$ K. The postshock temperature is scaled by the explosion energy
as $T=1.33\times
10^{10}\,\mathrm{K}(r_\mathrm{shock}/1000\,\mathrm{km})^{-3/4}(E_\mathrm{exp}/1\,\mathrm{Bethe})^{1/4}$,
where $r_\mathrm{shock}$ is the shock radius \citep{woos02}. Therefore
with $E_\mathrm{exp}=1\,\mathrm{Bethe}$, $^{56}$Ni can be generated
for $r_\mathrm{shock}\lesssim 3700$ km.  Since shock velocity $v_s$ is
roughly $10^4$ km s$^{-1}$ after the onset of the explosion, it takes
only a few hundred milliseconds to reach this radius.  If the growth
rate of the explosion energy is small and it takes a few second to
achieve 1 Bethe, it is not trivial whether $^{56}$Ni is synthesized by
explosive nucleosynthesis in the ejecta.

In this paper, we investigate $^{56}$Ni production as an indicator of
the explosion mechanism. First we perform numerical simulations of
supernova explosion in Section \ref{sec:simulation}. By calibrating
with numerical simulation data about shock and temperature evolution,
we construct an analytic model that describes shock and temperature
evolution, which are important ingredients of $^{56}$Ni production,
and give constraint on the growth rate of the explosion energy to
synthesize enough $^{56}$Ni in Section \ref{sec:analytic}. This
analytic model is useful to investigate $^{56}$Ni production for a
broader parameter space of both the explosion properties and
progenitor structure. We summarize our results and discuss their
implications in Section \ref{sec:summary}.

\begin{table}
\centering
\caption{Properties of recent explosion simulations}
\label{tab:models}
\begin{tabular}{lcc} 
\hline
Author(s) & ZAMS mass $^a$ & $\dot E_\mathrm{exp}$ $^b$\\
 & ($M_\odot$) & (Bethe s$^{-1}$)\\
\hline
2D (axisymmetric)\\
\hline
\cite{brue16} & 12, 15, 20, 25 & 1.5 -- 3 \\
\cite{suwa16a} & 12 -- 100 & 0.5 -- 0.7 \\
\cite{pan16} & 11, 15, 20, 21, 27 & 1 -- 5 \\
\cite{ocon15} & 12, 15, 20, 25 & 0.5 -- 1 \\
\cite{naka16} & 17 & 0.4 \\
\cite{summ16} & 11.2 -- 28 & 1 \\
\cite{burr16} & 12, 15, 20, 25 &  1 -- 3 \\
\hline
3D\\
\hline
\cite{lent15} & 15 & 0.2 \\
\cite{mels15a} & 9.6 & 0.6 \\
\cite{muel15b} & 11.2 & 0.4\\
\cite{taki16} & 11.2, 27 & 0.4 -- 2\\
\hline
\end{tabular}
\begin{flushleft}
$^a$ Not only the mass, evolution codes are also different.\\
$^b$ Note that these numbers are quite rough estimates in the early
  phase ($\sim 100$ ms after the onset of explosion) based on figures
  in the literature.
\end{flushleft}
\end{table}

\section{Numerical simulations}
\label{sec:simulation}

\subsection{Method}

We employ {\tt blcode}, which is a prototype code of {\tt SNEC}
\citep{moro15} and a pure hydrodynamics code\footnote{Both codes are
  available from https://stellarcollapse.org.} based on
\citet{mezz93}, as a base. It solves Newtonian hydrodynamics in
Lagrange coordinate. Basic equations are given by
\begin{align}
\frac{\partial r}{\partial M}&=\frac{1}{4\pi r^2\rho},\\
\frac{Dv}{Dt}&=-\frac{GM}{r^2}-4\pi r^2\frac{\partial P}{\partial M},\\
\frac{D\epsilon}{Dt}&=-P\frac{D}{Dt}\left(\frac{1}{\rho}\right)+\mathcal{H}-\mathcal{C},
\end{align}
where $r$ is radius, $M$ is mass coordinate, $\rho$ is density, $v$ is
radial velocity, $t$ is time, $G$ is the gravitational constant, $P$
is pressure, and $\epsilon$ is specific internal energy.  $D/Dt$ means
Lagrange derivative. Artificial viscosity by \cite{neum50} is employed
to capture a shock.  Neutrino heating and cooling are newly added in
this work by a method used in the literature \citep[e.g.][]{murp08},
in which neutrino cooling is given as a function of temperature and
neutrino heating is a function of radius with a parametric neutrino
luminosity.  Heating term, $\mathcal{H}$, and cooling term,
$\mathcal{C}$, are given as
\begin{align}
\mathcal{H}=&1.544\times 10^{20}\, \mathrm{erg\,g^{-1}\,s^{-1}}
\nonumber\\
&\times\left(\frac{L_{\nu_e}}{10^{52}\mathrm{erg \,s}^{-1}}\right)
\left(\frac{r}{100\mathrm{km}}\right)^{-2}
\left(\frac{T_{\nu_e}}{4\mathrm{MeV}}\right)^{2},\\
\mathcal{C}=&1.399\times 10^{20}\, \mathrm{erg\,g^{-1}\,s^{-1}}
\left(\frac{T}{2\mathrm{MeV}}\right)^{6}.
\end{align}
Here, we fixed neutrino temperature as $T_{\nu_e}=4$MeV. In addition,
we take into account these terms only in postshock regime. We do not
take into account optical depth terms \citep[see][]{nord10,hank12} for
simplicity.  We modify inner boundary conditions so that the innermost
mass element does not shrink within 50 km from the center to mimic the
existence of a protoneutron star.  The Helmholtz equation of state by
\cite{timm99} is used. Initial composition is used for equation of
state.

\begin{table*}
\centering
\caption{Precollapse properties of the SN progenitors from \citet{woos07}
 }
\label{tab:models2}
\begin{tabular}{lccccccccccccc} 
\hline
Name & ${M_{s=4}}^a$ &  ${R_{M_{s=4}}}^b$ & ${\rho_{M_{s=4}}}^c$ & ${\xi_{M_{s=4}}}^d$ & ${\mu_{M_{s=4}}}^e$ &  ${R_{{M_{s=4}}+0.1M_\odot}}^f$ & ${\rho_{M_{s=4}+0.1M_\odot}}^g$ & ${\xi_{M_{s=4}+0.1M_\odot}}^h$ & ${\mu_{M_{s=4}+0.1M_\odot}}^i$\\
  & ($M_\odot$) & (1000 km) & ($10^7$ g cm$^{-3}$) & & & (1000 km) & ($10^7$ g cm$^{-3}$) \\ 
\hline
WH07s12 &    1.530 &    2.813 &    0.168 &    0.544 &    0.084 &    4.655 &    0.035 &    0.350 &    0.048 \\ 
WH07s15 &    1.818 &    3.770 &    0.129 &    0.482 &    0.116 &    4.924 &    0.051 &    0.390 &    0.079 \\ 
WH07s20 &    1.824 &    2.654 &    0.268 &    0.687 &    0.119 &    3.646 &    0.133 &    0.528 &    0.112 \\ 
WH07s25 &    1.901 &    2.803 &    0.317 &    0.678 &    0.157 &    3.771 &    0.131 &    0.531 &    0.118 \\ 
\hline
\end{tabular}
\begin{flushleft}
$^a$ Mass with $s=4k_B$ baryon$^{-1}$.\\
$^b$ Radius with $s=4k_B$ baryon$^{-1}$.\\
$^c$ Density with $s=4k_B$ baryon$^{-1}$.\\
$^d$ Compactness parameter of ${M_{s=4}}$, see Eq. (\ref{eq:xiM}).\\
$^e$ $\mu$ parameter determined by Eq. (\ref{eq:muM}) in units of $M_\odot/1000$ km.\\
$^f$ Radius with ${M_{s=4}+0.1M_\odot}$.\\
$^g$ Density with ${M_{s=4}+0.1M_\odot}$.\\
$^h$ Compactness parameter of ${M_{s=4}}+0.1M_\odot$.\\
$^i$ $\mu$ parameter of ${M_{s=4}}+0.1M_\odot$ in units of $M_\odot/1000$ km.
\end{flushleft}
\end{table*}

The initial conditions are the 12, 15, 20, and 25 $M_\odot$ models
from \cite{woos07}.  Properties of the progenitor models are given in
Table \ref{tab:models2}. In this table, we show mass coordinate,
radius, and density at a mass coordinate which has $s=4k_B$
baryon$^{-1}$, since the current understanding of shock launch is that
it is realized when a mass element with $s=4 k_B$ baryon$^{-1}$ is
accreting onto the shock.  In the fifth column, we show the
``compactness parameter'' \citep{ocon11}, which is defined as
\begin{equation}
\xi_{M}=\frac{M/M_\odot}{R(M)/1000\,\mathrm{km}},
\label{eq:xiM}
\end{equation}
where $R(M)$ is the radius of the sphere whose mass coordinate is $M$.
According to \cite{ocon11}, smaller values of $\xi_M$ are better for
explosions, but note that they used $\xi_{2.5}$, which is different
from our values.
The sixth column gives $\mu_M$ \citep{ertl16}, which is defined as
\begin{equation}
\mu_{M}=\left.\frac{dM}{dr}\right|_{r=R(M)}=4\pi\rho R^2(M),
\label{eq:muM}
\end{equation}
in units of $M_\odot/1000$ km. Note that \cite{ertl16} evaluated the
value of $dM/dr$ by computing the numerical derivative at the mass
shell where $s=4k_B$ baryon$^{-1}$ with a mass interval of $0.3
M_\odot$. Here we instead simply use the second equality in equation
(\ref{eq:muM}) to compute $dM/dr$ analytically. They showed that for a
given value of $M_{s=4}$, a smaller $\mu_M$ is better for an
explosion. From seventh to tenth columns give the same quantities as
ones from third to sixth columns, but different mass coordinate
$M_{s=4}+0.1M_\odot$.

The mass cut is determined by $M_{s=4}-0.2M_\odot$.  We employ 1000
grids with mass resolution of $10^{-3}M_\odot$ so that $1M_\odot$ is
included in numerical regime. To check the impact of this choice, we
additionally perform a simulation with a mass cut of
$M_{s=4}-0.3M_\odot$ with 1100 grid points and find no significant
difference from standard grid model. We also performed a simulation
with 1500 grids points and with the same total mass (i.e. 33\% better
rezolution) and found no significant differences. Therefore, the
numerical results that will be shown below are insensitive to
numerical setup.

In the following, we use the so-called diagnostic explosion energy,
which is defined as the integral of the sum of specific internal,
kinetic, and gravitational energies over all zones, in which it is
positive, as an approximate estimate of the explosion energy. Note
that this energy is not direct observables, since there is bound
(totally negative energy) material above the shock and it reduces the
explosion energy when it is swept by the shock.

\subsection{Results}
\label{sec:result}

\begin{table*}
\centering
\caption{Summary of simulations}
\label{tab:simulation}
\begin{tabular}{lccccccccc} 
\hline \hline
Name & progenitor & ${L_{\nu_e,52}}^a$ & ${t_\mathrm{exp}}^b$ & ${t_{T_9=5}}^c$ & ${E_{\mathrm{exp},T_9=5}}^d$ & ${\dot{E}_\mathrm{exp,T_9=5}\,}^e$ & ${E_{\mathrm{exp},1\mathrm{s}}}^f$  & ${M_\mathrm{PNS}}^g$ & ${M_{^{56}Ni}}^h$\\
& & ($10^{52}$ erg s$^{-1}$)  & (ms) & (ms) & (Bethe) &(Bethe s$^{-1}$) & (Bethe) & ($M_\odot$) & ($M_\odot$)  \\
\hline
WH07s12L1 & WH07s12 & 1 &  --- &  --- & --- & --- & --- & --- & --- \\
WH07s12L2 & WH07s12 & 2 & 553 & 97 & 0.093 & 0.950 & 0.147 & 1.527 & 0.023 -- 0.047 \\ 
WH07s12L3 & WH07s12 & 3 & 361 & 130 & 0.230 & 1.769 & 0.478 & 1.456 & 0.068 -- 0.098 \\
WH07s12L4 & WH07s12 & 4 & 233 & 149 & 0.366 & 2.447 & 0.981 & 1.315 & 0.097 -- 0.226 \\
\hline
WH07s15L2 & WH07s15 & 2 & --- & --- & --- & --- & --- & --- & ---\\
WH07s15L3 & WH07s15 & 3 & 580 & 135 & 0.166 & 1.230 & 0.164 & 1.820 & 0.060 -- 0.079 \\ 
WH07s15L4 & WH07s15 & 4 & 409 & 151 & 0.358 & 2.362 & 0.502 & 1.737 & 0.086 -- 0.135 \\ 
WH07s15L5 & WH07s15 & 5 & 267 & 160 & 0.481 & 3.007 & 1.057 & 1.648 & 0.107 -- 0.196 \\
\hline
WH07s20L2 & WH07s20 & 2 & --- & --- & --- & --- & --- & ---\\ 
WH07s20L3 & WH07s20 & 3 & 307 & 197 & 0.344 & 1.752 & 0.575 & 1.806 & 0.118 -- 0.151 \\
WH07s20L4 & WH07s20 & 4 & 249 & 175 & 0.392 & 2.238 & 0.791 & 1.769 & 0.110 -- 0.166 \\ 
WH07s20L5 & WH07s20 & 5 & 236 & 169 & 0.458 & 2.709 & 1.042 & 1.736 & 0.107 -- 0.196 \\
\hline
WH07s25L2 & WH07s25 & 2 & --- & --- & --- & --- & --- & --- & ---\\ 
WH07s25L3 & WH07s25 & 3 & 427 & 210 & 0.379 & 1.801 & 0.591 & 1.943 & 0.125 -- 0.149 \\ 
WH07s25L4 & WH07s25 & 4 & 238 & 183 & 0.431 & 2.354 & 0.981 & 1.852 & 0.113 -- 0.172 \\
WH07s25L5 & WH07s25 & 5 & 226 & 171 & 0.492 & 2.874 & 1.220 & 1.822 & 0.111 -- 0.197 \\
\hline
\hline
\end{tabular}
\begin{flushleft}
$^a$ Neutrino luminosity.\\
$^b$ Time between NS formation and explosion onset.\\
$^c$ Time between explosion onset and postshock temperature being $T=5\times 10^9$ K.\\
$^d$ Explosion energy at a time when postshock temperature is $T=5\times 10^9$ K.\\
$^e$ $E_{\mathrm{exp},T_9=5}/t_{T_9=5}$.\\
$^f$ Explosion energy at 1 s after explosion onset.\\
$^g$ PNS mass at the last time of simulation.\\
$^h$ $^{56}$Ni mass.
\end{flushleft}
\end{table*}

The results are summarized in Table \ref{tab:simulation}.  Model names
are denoted as WH07sAALB, where the two digits AA indicate the
progenitor mass, and a digit B indicates the neutrino luminosity (see
the second and third columns in the same table).  WH07 means
\cite{woos07}.  $t_\mathrm{exp}$ is the explosion onset time (time at
the diagnostic explosion energy becoming positive) measured from
protoneutron star (PNS) formation time, which is determined by the
innermost mass element reaching $r=50$ km.  $t_{T_9=5}$ presents
post-explosion time when the temperature just after the shock becomes
$T_9=5$, where $T_9=T/10^9$ K, and the next column gives explosion
energy at the same time. $\dot{E}_\mathrm{exp,T_9=5}$ is the growth
rate of the explosion energy during
$t_{T_9=5}$. $E_{\mathrm{exp},1\mathrm{s}}$ is explosion energy at 1 s
after the explosion onset. $M_\mathrm{PNS}$ is final PNS mass which is
estimated by the locally bound material below shock wave. The last
column gives the mass of $^{56}$Ni which is calculated as the mass of
the material whose maximum temperature is over $5\times 10^9$K.  The
range implies the uncertainty in the simulation.  Since the PNS mass
(i.e. so-called mass cut in canonical nucleosynthesis studies) evolves
in time, we give minimum and maximum mass with maximum temperature
being beyond $5\times 10^9$K above PNS. The maximum value includes a
component ejected as a neutrino-driven wind. Whether this component
synthesizes $^{56}$Ni or not depends on the evolution of electron
fraction $Y_e$, which is altered by neutrino irradiation from
PNS. Since it is beyond the scope of this study, we do not discuss it
below.

\begin{figure}
\centering
\includegraphics[width=0.45\textwidth]{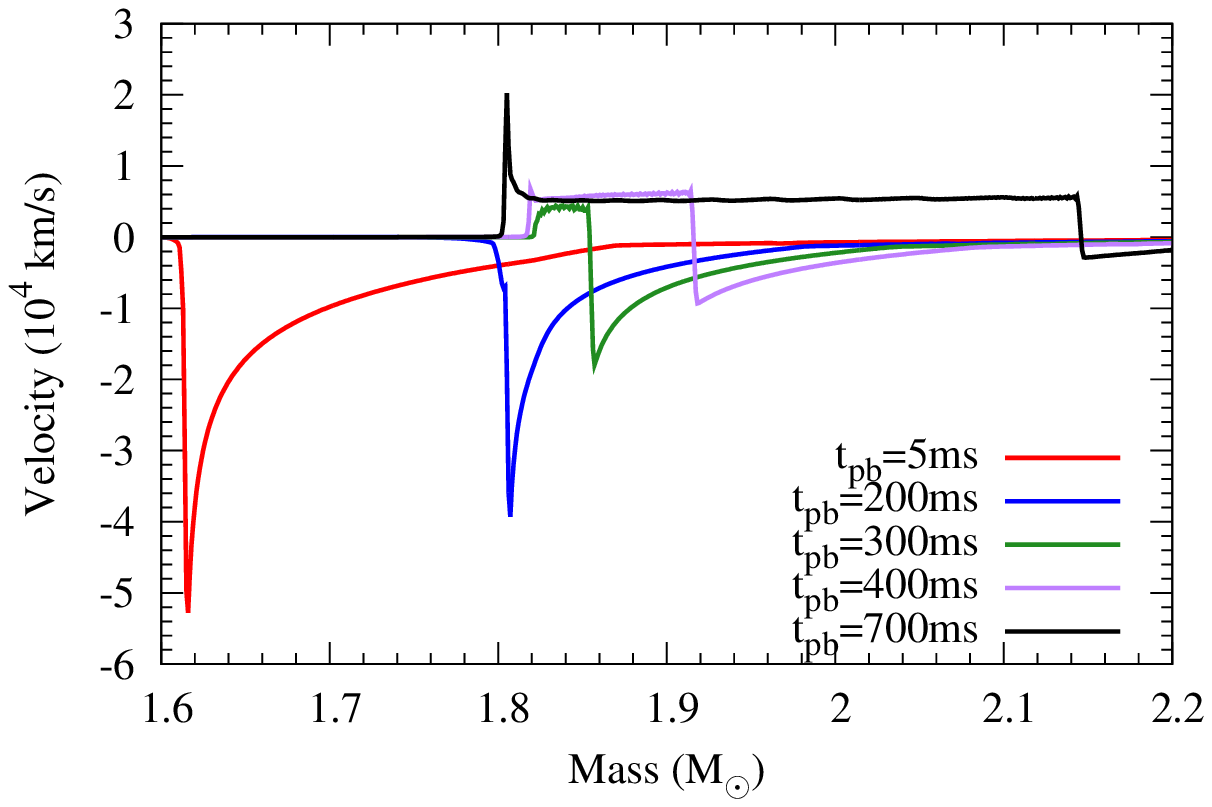}
\includegraphics[width=0.45\textwidth]{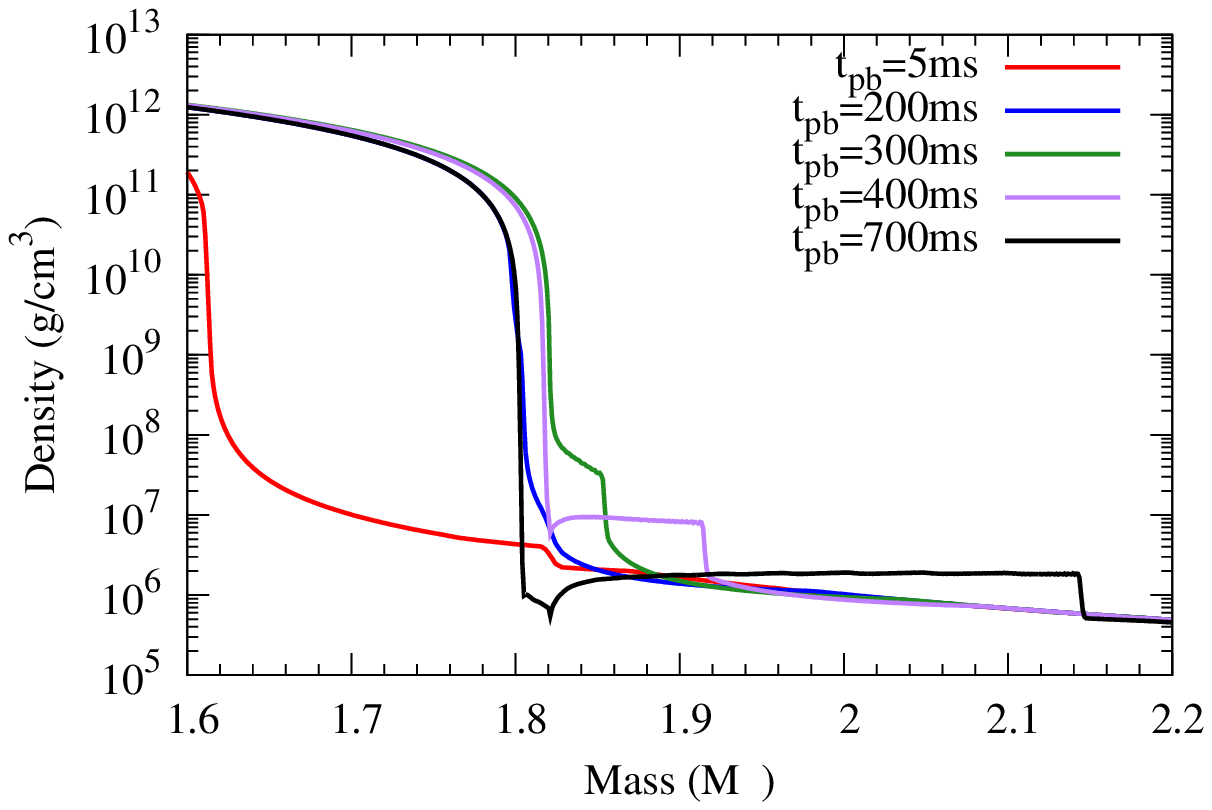}
\includegraphics[width=0.45\textwidth]{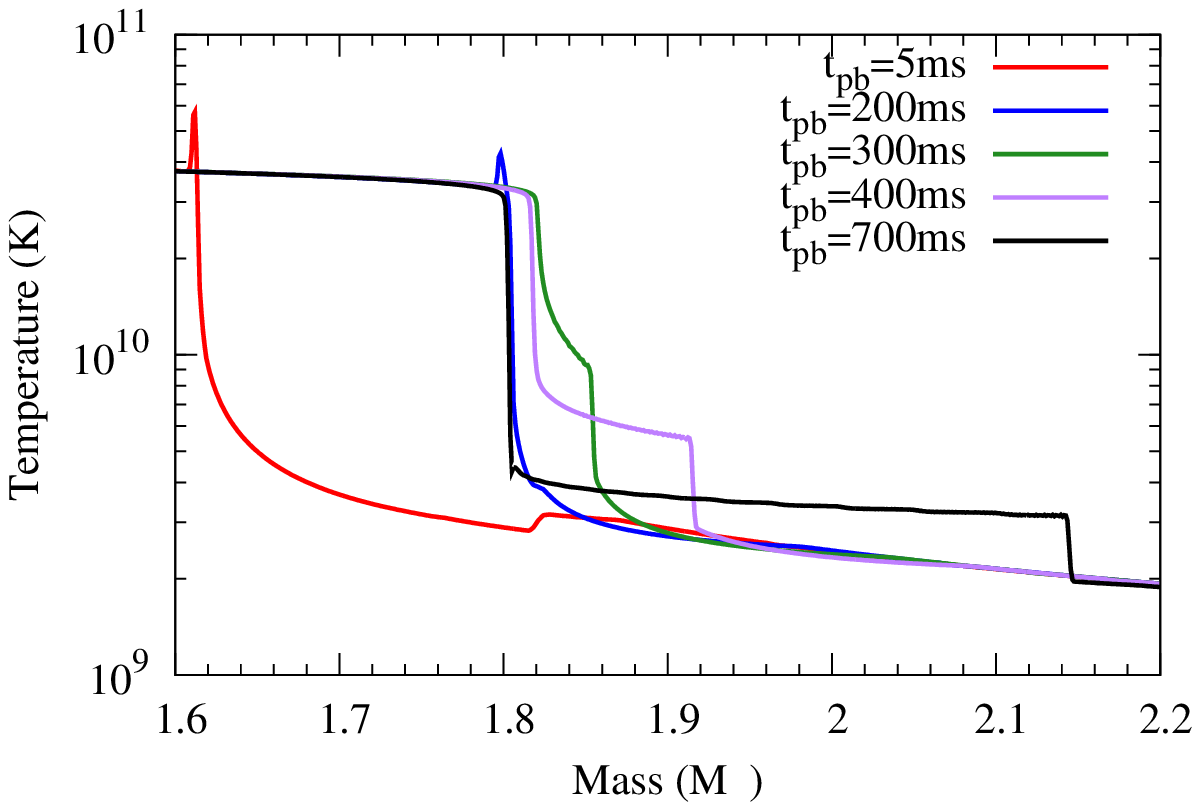}
\caption{ Time evolution of the radial velocity (top) density (middle)
  and temperature (bottom) as a function of mass coordinate for model
  WH07s20L4. Each line indicates different time from 5 ms to 700 ms
  after the bounce (i.e., postbounce time). The shock begins expansion
  at $t_\mathrm{pb}\sim 300$ ms.  }
\label{fig:m-vel}
\end{figure}

Figure \ref{fig:m-vel} presents time evolution of radial velocity,
density, and temperature as a function of mass coordinate for model
WH07s20L4. It is clearly shown that a stalled shock is formed at first
and then once the Si/O layer ($\approx 1.82M_\odot$) accretes onto the
shock, the shock eventually begins to propagate outward (indicated by
the positive post-shock velocity) because of the rapid decrease of the
ram pressure.

\begin{figure}
\includegraphics[width=0.5\textwidth]{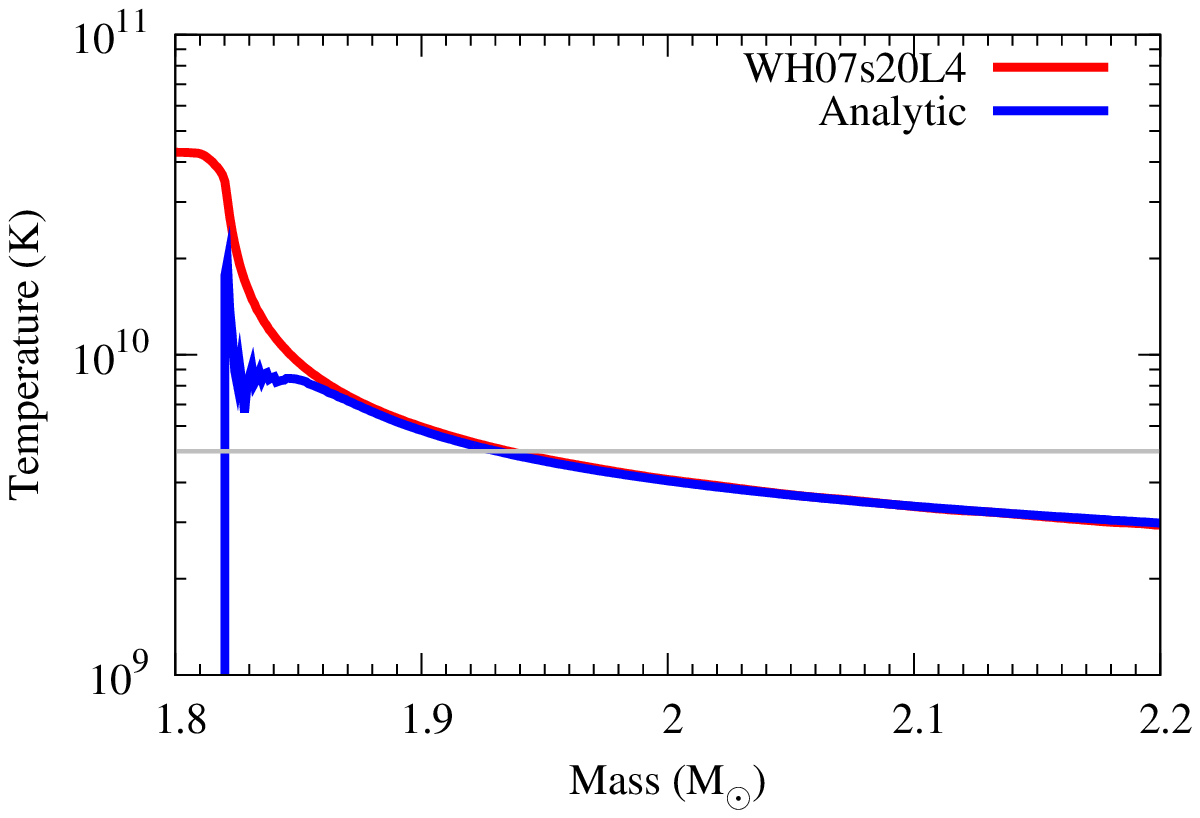}
\caption{Maximum temperature distributions of a numerical simulation
  (red line) and analytic expression (blue line).  Numerical model
  employs s20 model of \citet{woos07} and neutrino luminosity
  $L_\nu=4\times 10^{52}$ erg s$^{-1}$ and the consequent growth rate
  of the explosion energy is $\approx 2.2\times 10^{51}$ erg s$^{-1}$.
}
\label{fig:T(M)}
\end{figure}

Figure \ref{fig:T(M)} gives maximum temperature distribution as a
function of mass coordinate found in model WH07s20L4 (red line) and
analytic estimate based on the explosion energy (blue line). The
analytic estimate is given by solving the following equation:
\begin{align}
E_{\mathrm{exp}}=\frac{4\pi}{3}r_s^3 aT^4
f(T_9),
\end{align}
where $a=7.56\times 10^{-15}$ erg cm$^{-3}$ K$^{-4}$ is the radiation
constant and $r_s$ is the shock radius.  A temperature-dependent
function $f(T_9)=1+(7/4)T_9^2/(T_9^2+5.3)$ \citep{frei99,tomi09},
which takes into account both radiation and non-degenerate electron
and positron pairs, is used here. Since the temperature range is not
large, $f(T_9=5)=2.44$ also gives a rather good agreement with
numerical result. This factor makes the temperature smaller by 20\%
than one without the correction. In this estimate, the postshock
temperature in the ejecta is assumed to be a constant in space, which
is indeed realized in the simulation (see Figure \ref{fig:m-vel}). In
the analytic estimate shown in the figure, we take the explosion
energy and shock radius from the corresponding numerical simulation.

\begin{figure}
\includegraphics[width=0.5\textwidth]{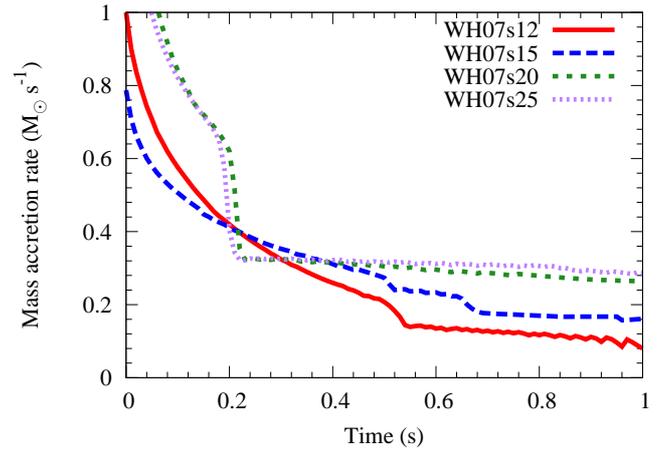}
\caption{ Mass accretion rate as a function of time after NS
  formation, measured at $r=500$ km. All models shown here are
  non-exploding models with a small neutrino luminosity.  }
\label{fig:mdot}
\end{figure}

Figure \ref{fig:mdot} presents time evolution of mass accretion rate
($\dot M$) of non-exploding models.  Mass accretion rates are measured
at $r=500$ km. Since there is a correlation between mass accretion
rate and explosion criteria via critical neutrino luminosity
\citep{burr93}, the mass accretion rate is a good measure to discuss
explodability. As is known, the mass accretion rate becomes almost
constant when Si/O layer is accreting \citep[see, e.g.][]{suwa16a},
which is seen in these simulation as well, especially in models
WH07s20 and WH07s25.  The constant values of accretion rate are
dependent on the initial progenitor structure. From table
\ref{tab:simulation}, one sees that models with high $L_{\nu_e}$
exhibit similar explosion onset time (4th column) for WH07s20 and
WH07s25, which have rapid transient in mass accretion rate (green and
purple lines). Meanwhile, WH07s12 and WH07s15 do not show such a clear
transition, i.e. the explosion onsets earlier for higher $L_{\nu_e}$,
since these progenitor do not have a drastic change of mass accretion
rate (red and blue lines).

\section{Analytic model}
\label{sec:analytic}

In this section, we derive the temperature evolution based on a simple
analytic model and justify it with numerical results explained in the
previous section.

\subsection{The expansion-wave collapse solution}

As known in star formation field, there is a self-similar solution of
stellar collapse, so-called ``expansion-wave collapse solution''
\citep{shu77}. This solution implies that the density structure inside
rarefaction wave becomes $\rho(r)\propto r^{-3/2}$ and $r^{-2}$
outside for {\it isothermal} gas. \cite{suto88} extended this solution
for adiabatic flow with arbitrary adiabatic index and showed that
$\rho\propto r^{-3/2}$ profile inside rarefaction wave is obtained
irrespective of adiabatic index.

From modern supernova simulations, typical progenitors lead to a
constant mass accretion rates when Si/O layer is accreting \citep[see
  Appendix A of][]{suwa16a}.  With this fact and continuity equation,
$\partial_t\rho+r^{-2}\partial_r(r^2\rho v)=0$, where
$\partial_t=\partial/\partial t$ and $\partial_r=\partial/\partial r$,
one recognizes that the density structure does not evolve,
i.e. $\partial_t\rho=0$, since $r^2\rho v=\dot M/4\pi$ becomes
constant.

The current understanding of explosion onset is the following. A rapid
density decrease between Si/O layers leads to decrease of the ram
pressure above the shock due to decreasing mass accretion rate. It
results in a shock expansion, since the thermal pressure changes more
slowly and overwhelms the ram pressure. Therefore, when the base of
the oxygen layer arrives at the shock, the shock expands and,
simultaneously, the density structure above shock becomes
quasi-stationary. In the following we neglect time evolution of
density structure above a shock wave.

\subsection{Shock wave evolution}
\label{sec:shock}

The shock velocity is given by Eq. (19) of \cite{matz99} as
\begin{equation}
v_s=0.794
\left(\frac{E_\mathrm{exp}}{M_{\rm ej}}\right)^{1/2}
\left(\frac{M_{\rm ej}}{\rho(r_s) r_s^3}\right)^{0.19},
\label{eq:vs}
\end{equation}
where $E_\mathrm{exp}$, $M_{\rm ej}$, and $r_s$ are explosion energy,
ejecta mass, and shock radius, respectively.  The ejecta mass is given
by
\begin{align}
M_{\rm ej}(t,r_s)=\dot M t+\int_{r_{\rm mc}}^{r_s} 4\pi r^2 \rho(r) dr,
\end{align}
where $r_{\rm mc}$ is the radius of mass cut, i.e. the initial
position of the shock. We assume the density profile as (see previous
subsection)
\begin{equation}
\rho(r)=\rho_R\left(\frac{r}{R}\right)^{-3/2},
\label{eq:rho(r)}
\end{equation} 
where $\rho_R$ and $R$ are constants. Adopting the mass accretion rate
as $\dot M=4\pi r_s^2\rho(r_s)v_{\rm acc}(r_s)= 2\pi \rho_R
R^{3/2}\sqrt{2GM}$ ($v_{\rm acc}=v_\mathrm{ff}/2=\sqrt{GM/2r_s}$,
where $v_\mathrm{ff}=\sqrt{2GM/r}$ is free-fall velocity, is used), we
get
\begin{align}
M_{\rm ej}(t,r_s)=2\pi\rho_R R^{3/2}
\left[
\sqrt{2GM}t+\frac{4}{3}\left(r_s^{3/2}-r_{\rm mc}^{3/2}\right)
\right].
\label{eq:Mej}
\end{align}
Here, we also assume a constant mass accretion rate. By assuming
$r_s=v_st$ with a constant shock velocity $v_s$, one finds that at the
early time ($t\lesssim
GM/v_s^3=0.19\,(M/1.4M_\odot)\,(v_s/10^4\,\mathrm{km\,s^{-1}})^{-3}$
s), a contribution from mass accretion (the first term in square
bracket of Eq. \ref{eq:Mej}) dominates the ejecta mass, and at the
late time the swept mass contribution (the second term in bracket)
dominates.  In the following we evaluate shock evolutions in two
extreme cases: i) ejecta mass is dominated by accreted mass and ii)
ejecta mass is dominated by swept mass.

\subsubsection{Accreted mass dominant case}

Let us assume that $M_\mathrm{ej}=\dot Mt$ by neglecting swept mass
contribution (second term in the square brackets of
Eq. \ref{eq:Mej}). We also assume a constant growth rate of the
explosion energy, $\dot E_\mathrm{exp}$, which gives
$E_\mathrm{exp}=\dot E_\mathrm{exp}t$, for simplicity.  Since
$v_s=dr_s/dt$, by introducing Eq. (\ref{eq:Mej}) to Eq. (\ref{eq:vs}),
we obtain the following time evolution of the shock:
\begin{align}
r_s(t)=\left(\frac{0.86 \dot E_\mathrm{exp}^{1/2}}{\dot M^{0.31} \rho_R^{0.19} R^{0.57/2}} t^{1.19}+r_\mathrm{mc}^{2.57/2}\right)^{2/2.57}.
\label{eq:rs}
\end{align}
Here we use an initial condition that $r_s(t=0)=r_\mathrm{mc}$. The
origin of time (i.e. $t=0$) is determined by the shock transition from
a steady-accretion shock to an expanding shock, i.e. the onset time of
the explosion. Here, we leave $\dot M$ as a free parameter because
$v_\mathrm{acc}$ is not always half free-fall velocity. This is
because a fluid element starts to fall down after the rarefaction
waves passes it and before that it stays in hydrostatic configuration
with no bulk velocity.  Accretion rate based on progenitor structure
will be given in Section \ref{sec:temp}.

\subsubsection{Swept mass dominant case}
\label{sec:swept}

Here we take into account swept mass contribution alone in
Eq. (\ref{eq:Mej}), which is
\begin{align}
M_\mathrm{ej}(t,r_s)=\frac{8\pi}{3}\rho_R R^{3/2}(r_s^{3/2}-r_\mathrm{mc}^{3/2}).
\end{align}
Assuming $r_s\gg r_\mathrm{mc}$ and taking the leading order term of
$(r_\mathrm{mc}/r_s)$, we can integrate Eq. (\ref{eq:vs}) as
\begin{align}
&\left[\frac{4}{7}-1.24\left(\frac{r_\mathrm{mc}}{r_s}\right)^{3/2}\right]r_s^{7/4}-0.669r_\mathrm{mc}^{7/4}\nonumber\\
&\,\,\,\,\,\,=0.274\rho_R^{-1/2}R^{-3/4}\dot E_\mathrm{exp}^{1/2}t^{3/2},
\label{eq:swept}
\end{align}
where we imposed an initial condition of $r=r_\mathrm{mc}$ for
$t=0$. We can get shock evolution by solving this algebraic equation
numerically.

By comparing Eq. (\ref{eq:rs}) and solution of Eq. (\ref{eq:swept})
with a direct integrated solution of Eq. (\ref{eq:vs}), we find that,
in the parameter regime we are interested in, shock evolution is well
captured by these approximate solutions (i.e. Eqs. \ref{eq:rs} and
\ref{eq:swept}). For instance, with $\rho_R=10^7$ g cm$^{-3}$,
$R=1000$ km, $\dot M=0.5M_\odot$ s$^{-1}$, and $\dot E_\mathrm{exp}=1$
Bethe s$^{-1}$, differences between these three solutions keep within
$\sim$ 20\% for $r_s\lesssim 10,000$ km. Therefore, in Section
\ref{sec:temp} we use Eq. (\ref{eq:rs}) to estimate temperature
evolution, since this solution can be written in simply analytic
manner.

\subsection{Mass of ejecta}

In the above estimates, we assumed that all shocked materials which
accrete or are swept are included in ejecta mass. This assumption is
not always correct, since part of them accrete onto a neutron star
when postshock velocity is not outgoing. From Rankine-Hugoniot
condition, we have following relation at shock frame:
\begin{align}
\rho_\mathrm{pre} v_\mathrm{pre}=\rho_\mathrm{post} v_\mathrm{post},
\end{align}
where quantities with ``pre'' mean pre-shock states and ``post'' mean post-shock states.
By going to rest frame, we have
\begin{align}
\rho_\mathrm{pre} (v_\mathrm{pre}-v_s)=\rho_\mathrm{post} (v_\mathrm{post}-v_s).
\end{align}
The preshock and postshock densities are related by
$\rho_\mathrm{post}=\beta\rho_\mathrm{pre}$, where $\beta\approx 4$
\citep{muel16}.\footnote{This value is different from a strong shock
  limit, $\rho_\mathrm{post}/\rho_\mathrm{pre}=7$, for $\gamma=4/3$,
  where $\gamma$ is adiabatic index. This is because the Mach number
  of preshocked accretion flow is $\mathcal{M}\approx 3$
  \citep{muel98}, which gives
  $\rho_\mathrm{post}/\rho_\mathrm{pre}=4.2$ \citep[see Eq. 56.41
    of][]{miha84}.} In order to make postshock velocity positive
(i.e. $v_\mathrm{post}>0$), $v_s>-v_\mathrm{pre}/(\beta-1)\approx
-v_\mathrm{pre}/3$. Note that preshock velocity is negative
($v_\mathrm{pre}<0$), i.e. accreting, here.  It should be noted that
the same constraint is obtained even when we additionally employ
momentum conservation equation.  By combining this requirement with
Eq. (\ref{eq:vs}), we can estimate the ejecta mass excluding infalling
material at the onset of the explosion as follows.
\begin{align}
M_\mathrm{ej}(t\approx 0,r_s)&=\left[3.36(GM)^{-1/2}E_\mathrm{exp}^{1/2}(\rho_RR^{3/2})^{-0.19}r_s^{0.215}\right]^{1/0.31}\\
&=0.019M_\odot M_{1.4}^{-1.61}E_{\mathrm{exp},49}^{1.61}\rho_{R,7}^{-0.613}R_8^{-0.919}r_{s,7}^{0.694},
\end{align}
where $M_{1.4}=M/1.4M_\odot$, $E_{\mathrm{exp},49}=E_{\mathrm{exp}}/10^{49}$ erg, $\rho_{R,7}=\rho_R/10^7$ g cm$^{-3}$, $R_8=R/10^8$ cm, and $r_{s,7}=r_s/10^7$ cm.
Here we assume $|v_\mathrm{pre}|=\sqrt{GM/2r_s}$ and
$\rho(r_s)=\rho_R(R/r_s)^{3/2}$. This equation implies that ejecta
mass at the very beginning of the explosion ($E_\mathrm{exp}=10^{49}$
erg in this estimate) is negligible. For a case with a slow growth of
the explosion energy, i.e. a small $\dot E_\mathrm{exp}$, ejecta mass
should keep small and most of mass, which accretes onto the shock or
swept by the shock, must go through the ejecta and accrete onto a
central object (a neutron star or a black hole).

Note that for large $\dot E_\mathrm{exp}$ cases, a shock is rapidly
accelerated and accreting and swept materials are following the shock
as ejecta. Therefore, assumption employed in the previous subsection
is validated.

\subsection{Critical neutrino luminosity and heating rate}

In this subsection, we derive a critical value of the heating rate to
produce the explosion, based on discussion of a critical neutrino
luminosity in the literature. Below this critical value, the shock
cannot be launched.

It is well known that there is a critical neutrino luminosity to
produce an explosion driven by neutrino heating
mechanism. \cite{burr93} indicated a critical neutrino luminosity as a
function of mass accretion rate as $L_{\nu_e}\propto \dot M^{1/2.3}$,
in which neutrino average energy was assumed to be a constant. More
recently, subsequent studies updated the expression of critical
neutrino luminosity by taking into account other physical parameters,
e.g. neutrino average energy, PNS radius, etc. Here, we utilize
\cite{jank12}, which gives $L_{\nu,c}(\dot M)\propto\dot
M^{2/5}M_\mathrm{NS}^{4/5}$.
In the current set up, we found that the critical neutrino luminosity
for s20 is $L_{\nu_e,c}\approx 2.7\times 10^{52}$ erg s$^{-1}$, with
$M_\mathrm{NS}\approx 1.8M_\odot$ (determined by $M_{s=4}$) and $\dot
M\approx 0.3\,M_\odot$ s$^{-1}$. By changing parameters to
$M_\mathrm{NS}\approx 1.5M_\odot$ and $\dot M\approx 0.15\,M_\odot$
s$^{-1}$, which are relevant for s12, we get $L_{\nu_e,c}\approx
1.8\times 10^{52}$ erg s$^{-1}$. For s15, i.e. $M_\mathrm{NS}\approx
1.8M_\odot$ and $\dot M\approx 0.2\,M_\odot$ s$^{-1}$,
$L_{\nu_e,c}\approx 2.3\times 10^{52}$ erg s$^{-1}$.  These values are
roughly consistent with our numerical results.  Since the mass
accretion rate evolution is not constant in s12 and s15, the direct
comparison is not very meaningful. We do not try to derive more
precise estimate, because it is not the main focus of this study.

The heating rate by neutrino can be estimated with Eqs. (83) and (86)
of \cite{jank01} as
\begin{align}
\dot E_\mathrm{exp}=\mathcal{H-C}=3.3\times 10^{50}\mathrm{erg\,s^{-1}}
(2L_{\nu_e,52})\rho_{s,9}r_{s,7}(r_s/r_g)^2,
\label{eq:Edot}
\end{align}
where $r_g$ is the gain radius, which is $\approx 100$ km, and
$\rho_{s,9}$ is density behind the shock in units of $10^9$ g
cm$^{-3}$.  With $\dot M=4\pi r^2\rho v$ and compression ratio
$\beta=4$, $\rho_{s,9}=0.1 4\dot
M_{0.3}M_\mathrm{NS,1.8}^{-1/2}(r_{s,7}/2)^{-3/2}$, where $\dot
M_{0.3}=\dot M/0.3M_\odot\,\mathrm{s}^{-1}$,
$M_\mathrm{NS,1.8}=M_\mathrm{NS}/1.8M_\odot$, which are relevant for
WH07s20. $v=\sqrt{GM_\mathrm{NS}/2r_s}$ is again used.  For
$L_{\nu_e,52}=4$, $r_{s,7}=2$, and $r_s/r_g=2$, we get $\dot
E_\mathrm{exp}=2.7\times 10^{51}$ erg s$^{-1}$, which roughly agrees
with model WH07s20L4 (see Table \ref{tab:models2}). By combining
critical neutrino luminosity, we can derive a critical heating rate to
produce an explosion as
\begin{align}
\dot E_{\mathrm{exp},c}=1.9\times 10^{51}\mathrm{erg\,s^{-1}}
\dot M_{0.3}^{7/5} M_{1.8}^{3/10}(r_{s,7}/2)^{1/2}(r_s/2r_g).
\label{eq:Edot_crit}
\end{align}
This is slightly larger than a consequent value of model WH07s20L3
($\dot E_\mathrm{exp}=1.8\times 10^{51}$ erg s$^{-1}$). The
inconsistency is originated from the simplification of the analytic
model, which we do not discuss further.

\subsection{Temperature evolution}
\label{sec:temp}

The temperature can be estimated by
\begin{equation}
\frac{4\pi}{3} r_s^3 a T^4\zeta=E_\mathrm{int}+\dot E_\mathrm{exp}t,
\end{equation}
where $E_\mathrm{int}$ is the initial internal energy and $\zeta=2.44$
(see Section \ref{sec:result}).  Here we assume that $E_\mathrm{int}$
is compensating for gravitational binding energy at onset of the
explosion (i.e. the explosion energy becomes positive) so that it does
not appear in the expression of the explosion energy.  From this
equation, the temperature is written as
\begin{align}
T
&=\left(\frac{3(E_\mathrm{int}+\dot E_\mathrm{exp}t)}{4\pi r_s^3 a\zeta}\right)^{1/4}\\
&=6.0\times 10^{10}\,\mathrm{K}(E_{\mathrm{int},51}+\dot E_\mathrm{exp,51}t_0)^{1/4}r_{s,7}^{-3/4},
\label{eq:T}
\end{align}
where $E_{\mathrm{int},51}=E_\mathrm{int}/10^{51}$ erg and $\dot
E_\mathrm{exp,51}=\dot E_\mathrm{exp}/10^{51}$ erg s$^{-1}$. By
combining Eqs. (\ref{eq:rs}) and (\ref{eq:T}), we get
\begin{align}
T=&6.0\times 10^{10}\,\mathrm{K}(E_{\mathrm{int},51}+\dot E_\mathrm{exp,51}t_0)^{1/4}\nonumber\\
& \times\left(\frac{320\dot E_\mathrm{exp,51}^{1/2}}{\dot M_0^{0.31} \rho_{R,7}^{0.19} R_8^{0.57/2}} t_0^{1.19}+r_\mathrm{mc,7}^{2.57/2}\right)^{-3/5.14},
\label{eq:Tnum}
\end{align}
where $\dot M_0=\dot M/M_\odot$ s$^{-1}$.

Next, we derive $E_\mathrm{int}$ that dominates the temperature
evolution in the early phase, from stellar structure. Since a standing
accretion shock turns to a runaway phase when the thermal pressure in
postshock regime becomes larger than the ram pressure in preshock
regime, the time evolution of ram pressure is crucial. The preshock
ram pressure can be evaluated by the free-fall model as
\begin{equation}
P_{\mathrm{ram}}=\rho v_\mathrm{acc}^2=\frac{\dot M_s}{4\pi r_s^2}v_\mathrm{acc},
\end{equation}
where $M_s$ is a total mass enclosed by the shock and $\dot M_s$ is
mass accretion rate at the shock. Here we assume that $M_s+\dot
M_s\delta t\approx M_s$, where $\delta t$ is the timescale we are
interested in.  The mass accretion rate is \citep{woos12}
\begin{equation}
\dot M_s=\frac{dM_s}{dt_\mathrm{ff}}
=
\frac{2M_s}{t_\mathrm{ff}}\frac{\rho_0}{\bar\rho_0-\rho_0},
\end{equation}
where $\rho_0$ is the density at $t=0$ and $\bar\rho_0=3M_s/(4\pi
r_0^3)$ is the mean density inside $r_0$ (initial radius of the mass
shell). $t_\mathrm{ff}$ is the free-fall time, which is
$t_\mathrm{ff}=\sqrt{3\pi/(32G\bar\rho_0)}=\sqrt{\pi^2r_0^3/(8GM_s)}$
\citep{kipp90}. By combining them and using $\bar\rho_0\gg\rho_0$, we
get
\begin{equation}
P_\mathrm{ram}=\frac{4}{3\pi}\frac{GM_s}{r_0}\rho_0\left(\frac{r_0}{r_s}\right)^{5/2}.
\end{equation}
Therefore, the internal energy of postshock regime is given by
\begin{equation}
e_\mathrm{int}=3P_\mathrm{ram}=\frac{4}{\pi}\frac{GM_s}{r_0}\rho_0\left(\frac{r_0}{r_s}\right)^{5/2}.
\end{equation}
Here we assume that the pressure is dominated by radiation component, i.e. $\gamma=4/3$.

Then, $E_\mathrm{int}$ can be estimated as
\begin{align}
E_\mathrm{int}&=\frac{4\pi r_s^3}{3}\times 3P_\mathrm{rad}\\
&=\frac{16}{3}GM_s\rho_0 r_0^{3/2}r_s^{1/2}\\
&=3.13\times 10^{49}
\left(\frac{M_s}{1.4M_\odot}\right)
\left(\frac{\rho_0}{10^{7}\,\mathrm{g\,cm^{-3}}}\right)\nonumber\\
&\,\,\times\left(\frac{r_s}{100\,\mathrm{km}}\right)^{1/2}
\left(\frac{r_0}{1000\,\mathrm{km}}\right)^{3/2}
\mathrm{erg}.
\label{eq:Eint}
\end{align}
Note that this value is {\it not} an actual total internal energy
included by the shock, but is a rough estimate of an initial internal
energy of the ejecta which consists of a thin shell that is promptly
exploding.

\begin{figure}
\includegraphics[width=0.5\textwidth]{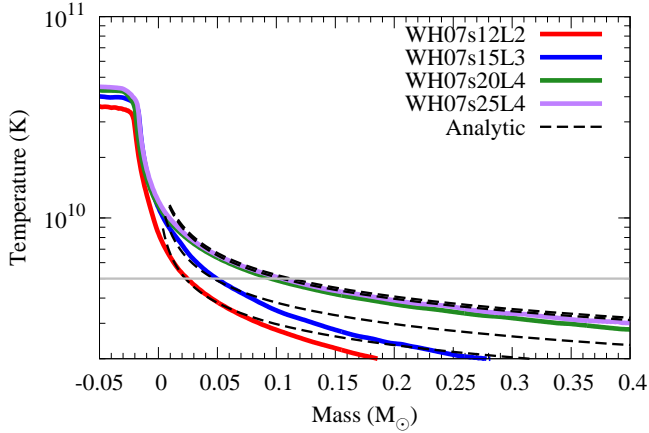}
\caption{ Maximum temperature distributions of four numerical
  simulations (colored solid lines) and analytic expression (black
  dashed lines). For analytic models, we use $\dot E_\mathrm{exp}$
  taken from Table \ref{tab:simulation}, $\rho$ and $R$ of
  $M_{s=4}+0.1M_\odot$ which are taken from Table \ref{tab:models2},
  origin of mass coordinate set to $M_{s=4}$, and
  $r_{\mathrm{mc},7}=2$.  $\dot M$ in analytic models are 0.15
  (WH07s12), 0.2 (WH07s15), 0.3 (WH07s20), and 0.3 $M_\odot$ s$^{-1}$
  (WH07s25), respectively, which are taken from Figure
  \ref{fig:mdot}. Numerical results are horizontally sifted by
  $0.02M_\odot$ (WH07s12, WH07s20, and WH07s25) and $0.03M_\odot$
  (WH07s15) leftward for direct comparison with analytic lines.  }
\label{fig:maxtemp}
\end{figure}

In Figure \ref{fig:maxtemp}, we show a comparison between numerical
results and analytic solutions for the maximum temperature
distribution as a function of mass coordinate. We pick up WH07s12L2,
WH07s15L3, WH07s20L4, and WH07s25L4, for typical models, since these
models start exploding when the mass accretion rate is (almost)
constant (see Table \ref{tab:simulation} and Figure \ref{fig:mdot}).
For analytic models, we solve shock evolution by Eq. (\ref{eq:rs}),
which only includes accreted mass in the ejecta mass, but we also add
swept mass by using Eq. (\ref{eq:rho(r)}) and values ($\rho$ and $R$)
at $M=M_{s=4}+0.1M_\odot$ from Table \ref{tab:models2}.  This
approximation works well, since the shock evolution by
Eq. (\ref{eq:rs}) is not largely different from a direct numerical
integration of Eq. (\ref{eq:vs}) (see Section \ref{sec:swept}). In
addition, we use $\dot E_\mathrm{exp}$ taken from Table
\ref{tab:simulation}, origin of mass coordinate set to $M_{s=4}$, and
$r_{\mathrm{mc},7}=2$. $\dot M$ in analytic models are 0.15 (WH07s12),
0.2 (WH07s15), 0.3 (WH07s20), and 0.3 (WH07s25), respectively, which
are taken from Figure \ref{fig:mdot}. Numerical results are
horizontally shifted by $0.02M_\odot$ (WH07s12, WH07s20, and WH07s25)
and $0.03M_\odot$ (WH07s15) leftward for direct comparison with
analytic lines in Figure \ref{fig:maxtemp}.  These shifts are showing
systematic error in analytic models, but these error is small enough
to discuss conventional amount of $^{56}$Ni, i.e. 0.07$M_\odot$ (for
SN 1987A, 1993J, and 1994I).  Numerical and analytic models of
WH07s20L4 and WH07s25L4 agree rather well for most regime, since these
models have considerably constant mass accretion rate (see Figure
\ref{fig:mdot}). On the other hand, WH07s12L2 and WH07s15L3 show
deviation between numerical and analytic models, especially in the
late time (i.e. large mass coordinate), because these models have
evolving mass accretion rates that break our assumption. Nevertheless,
temperature profile where we are interested in, i.e. $T_9>5$, are well
reproduced by the analytic models.

\subsection{Multidimensional effects}
\label{sec:multi-D}

 \begin{figure}
\includegraphics[width=0.5\textwidth]{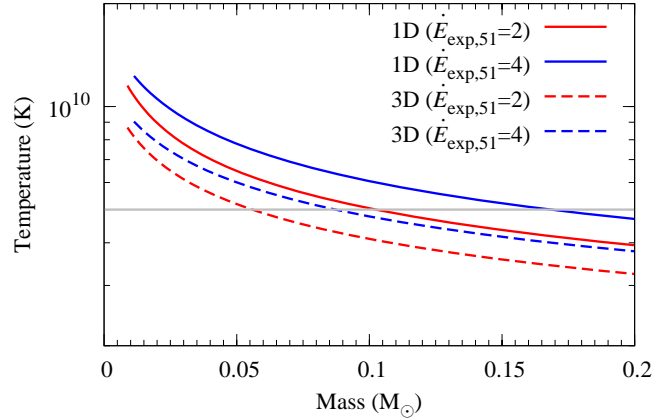}
\caption{The same plot as Figure \ref{fig:maxtemp}, but only analytic
  solutions shown for the model WH07s20. Solid and dashed lines
  indicate one-dimensional (1D) evolution and three-dimensional (3D)
  ones, respectively. Red and blue lines indicate different growth
  rates of the explosion energy, $\dot E_\mathrm{exp}$,
  respectively. A critical temperature for $^{56}$Ni production
  ($T=5\times 10^9$ K) is also presented by grey horizontal line.  }
\label{fig:M-Tmax}
\end{figure}

Next, let us introduce multidimensional (multi-D) effects in the
analytic model. It turns out from recent neutrino-radiation
hydrodynamics simulations that postshock pressure is not determined by
thermal pressure alone, but turbulent pressure (i.e. Reynolds stress)
is also contributing.  Roughly speaking, the turbulent pressure
becomes comparable to the thermal pressure \citep[e.g.][]{couc15a}. In
addition, at the propagating phase the kinetic energy becomes
comparable to the internal energy in the ejecta \citep[see,
  e.g. Figure 14 in ][]{brue16}. Therefore, it is natural to introduce
a factor ($\approx 0.5$) for internal energy amount in
Eqs. (\ref{eq:Tnum}) and (\ref{eq:Eint}), to take into account multi-D
effects, i.e.
\begin{align}
E_\mathrm{int}+\dot E_\mathrm{exp}t\to\frac{1}{2}\left(E_\mathrm{int}+\dot E_\mathrm{exp}t\right).
\label{eq:reduction}
\end{align}
Figure \ref{fig:M-Tmax} shows the impact of multi-D effect on the
temperature evolution. As is shown, the temperature of multi-D model
decreases compared to one-dimensional model. We also represent the
dependence of $\dot E_\mathrm{exp}$ in this figure. Roughly speaking,
multi-D models produce half amount of $^{56}$Ni of one-dimensional
models, which is consistent with consequence of \cite{yama13}, in
which they performed hydrodynamics simulations as well as
nucleosynthesis calculations of 1D and 2D (axial symmetry).

Even below the critical heating rate derived for the 1D cases,
successful explosions were observed in multi-D simulations.  Multi-D
effect is not only reducing the internal energy as explained above,
but also reducing critical neutrino luminosity
\citep[e.g.][]{murp08,hank12,couc13b}. Previous works typically showed
that multi-D simulations imply a smaller critical neutrino luminosity
for the explosion than 1D ones by $\sim 20$\%, depending on progenitor
model. From Eq. (\ref{eq:Edot}), the critical $\dot E_\mathrm{exp}$ is
proportional to $L_{\nu_e}$, the critical heating rate would be also
reduced by $\sim 20$\% in multi-D simulations.
In addition, multi-D simulations would produce partial explosions. In
particular, it is often seen in two-dimensional simulations that a
part of material explodes (polar direction) and other part forms a
downflow accreting onto a PNS. These structure reduces both diagnostic
explosion energy and ejecta mass, and leads to smaller amount of
$^{56}$Ni. We employ the following expression to take into account
partial explosion effect on the amount of $^{56}$Ni;
\begin{align}
M_{^{56}\mathrm{Ni}}=M_{^{56}\mathrm{Ni},c}\frac{\dot E_\mathrm{exp}}{\dot E_\mathrm{exp,c}},
\label{eq:M56Ni}
\end{align}
where $M_{^{56}\mathrm{Ni},c}$ is the amount of $^{56}$Ni
corresponding to critical heating rate in multi-D model.  It is worthy
to note that spherical symmetric explosion maximizes the amount of
$^{56}$Ni \citep{maed09,suwa15a}.

\subsection{Ejected $^{56}$Ni mass}
\label{sec:56ni}
 
 \begin{figure}
\includegraphics[width=0.5\textwidth]{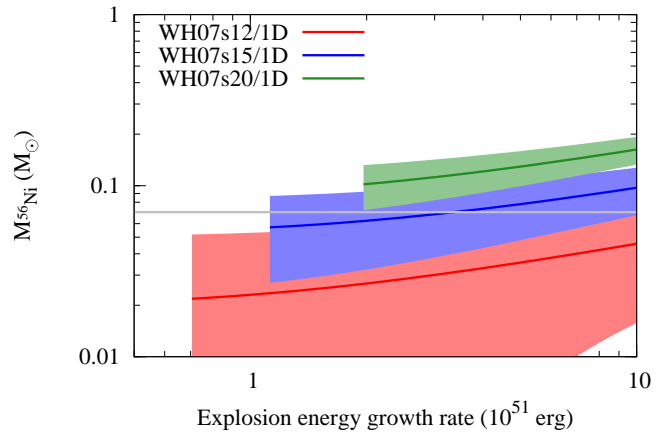}
\caption{The amount of $^{56}$Ni as a function of the growth rate of
  the explosion energy, $\dot E_\mathrm{exp}$.  Horizontal grey line
  indicates a canonical value of $^{56}$Ni, $0.07M_\odot$. Thick lines
  give analytic estimate with the same parameter sets as Figure
  \ref{fig:maxtemp} but different $\dot E_\mathrm{exp}$. Colored
  regions present possible error with $\pm 0.03M_\odot$, which is
  caused by, for instance, neutrino-driven wind upwards or fallback
  downward. The left endpoints correspond to the critical $\dot
  E_\mathrm{exp}$, which are estimated by
  Eq. (\ref{eq:Edot_crit}). Since WH07s25 indicate rather similar
  result as WH07s20 (see Figure \ref{fig:maxtemp}), it is not shown in
  this figure.}
\label{fig:L-Mni}
\end{figure}

In this subsection, we explain the amount of $^{56}$Ni depending on
the explosion energy growth rate and progenitor models.
Figure \ref{fig:L-Mni} presents the amount of $^{56}$Ni as a function
of $\dot E_\mathrm{exp}$ in 1D cases. All parameters other than $\dot
E_\mathrm{exp}$ are the same as Figure \ref{fig:maxtemp}. Thick lines
give analytic estimate and colored region show uncertainty of
models. For instance, neutrino-driven wind increases the amount of
$^{56}$Ni, definitely dependent on $Y_e$ profile of the wind, and
fallback of ejecta conversely decreases $^{56}$Ni. Since the impact of
these effects is largely uncertain, we here roughly present error
region with $\pm 0.03M_\odot$ as a guideline.
It should be noted that this figure implies discrepancy between our
numerical models and analytic model, especially for WH07s12 and
WH07s15 with a rather larger $\dot E_\mathrm{exp}$ than critical
value, since these models show time-evolving mass accretion rate,
which breaks the assumption employed in the analytic model. The
numerical models, however, employ a constant neutrino luminosity,
which means feedback effects of mass accretion rate evolution are
neglected. A natural expectation of the feedback effect is that the
neutrino luminosity decreases as the mass accretion rate
decreases. Then, shock launch is obtained once the mass accretion rate
reaches a stationary state with a constant mass accretion rate, which
exists for WH07s12 and WH07s15 as well, but rather late time (see
Figure \ref{fig:mdot}). Therefore, our analytic model works well.

\begin{figure}
\includegraphics[width=0.5\textwidth]{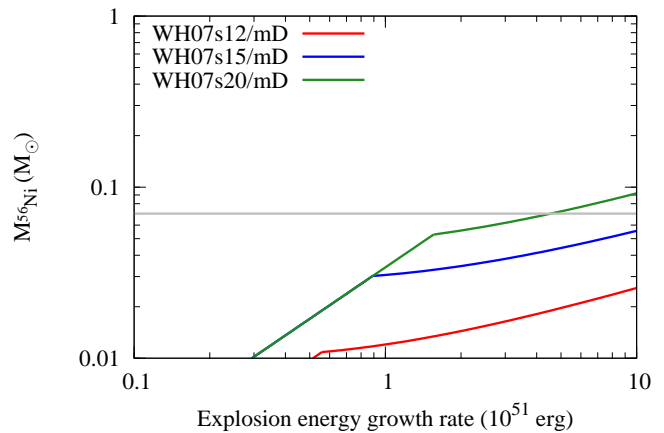}
\caption{The same plot as Figure \ref{fig:L-Mni}, but for
  multi-dimensional cases, in which reduction of thermal energy
  (Eq. \ref{eq:reduction}), reduction of the critical heating rate (by
  20\% from Figure \ref{fig:L-Mni}), and reduction of ejecta mass
  (Eq. \ref{eq:M56Ni}) are taken into account. The reduction of ejecta
  mass is only taken into account below the critical heating rate,
  which makes bend of lines around $\dot E_\mathrm{exp,51}\approx 1$.
}
\label{fig:L-Mni_mD}
\end{figure}

In Figure \ref{fig:L-Mni_mD}, we show the amount of $^{56}$Ni by
multi-D cases, in which reduction of thermal energy
(Eq. \ref{eq:reduction}), reduction of critical heating rate (by 20\%
from 1D) and reduction of ejecta mass (Eq. \ref{eq:M56Ni}) are all
taken into account.
As is shown, to achieve enough $^{56}$Ni synthesis, we need rather
large growth rate of the explosion energy, larger than $\approx$4
Bethe s$^{-1}$ for WH07s20 and even larger for WH07s12 and
WH07s15. Note that in this estimate, we do not include contribution
from neutrino-drive wind which is largely uncertain in this study.
\cite{brue16} indicated the amount of ejected $^{56}$Ni, in which both
explosive nucleosynthesis component and neutrino-driven wind component
are included, as 0.035 (WH07s12), 0.077 (WH07s15), 0.065 (WH07s20),
and 0.074 (WH07s20) $M_\odot$, respectively. The growth rate of the
explosion energy is roughly, $\sim 1.5$ (WH07s12), $\sim 2$ (WH07s15),
$\sim 2.5$ (WH07s20), and $\sim 3$ (WH07s25) Bethe s$^{-1}$,
respectively. Therefore, by taking contributions of explosive nuclear
burning from our analytic model, we find that neutrino-driven wind
contributes for $\sim 0.02$ (WH07s12), $\sim 0.04$ (WH07s15), $\sim
0.01M_\odot$ (WH07s20 and WH07s25), respectively. It is worthy to note
that their simulations in 2D exceptionally succeeded to produce enough
$^{56}$Ni, but their 3D model \citep{lent15} exhibited a much smaller
$\dot E_\mathrm{exp}$ than 2D (see Table \ref{tab:models}), which
implies difficulty of $^{56}$Ni synthesis in their 3D simulation.

\section{Summary and discussion}
\label{sec:summary}

$^{56}$Ni is an important indicator of the supernova explosion, which
characterizes light curves, particularly late decay phase. In
principle, the amount of $^{56}$Ni can be directly measured by light
curve alone, while ejecta mass and explosion energy are estimated by
combining light curve and spectrum properties. Nevertheless, the
explosion energy has often been paid attention from explosion
mechanism community, since it is easier to estimate from numerical
data than the amount of $^{56}$Ni. The final explosion energy,
however, is difficult to estimate by detailed numerical simulations,
which solve hydrodynamics equations as well as neutrino-radiation
transfer equation. This is because current simulations can reach only
$\mathcal{O}(1)$ s, but the explosion energy can grow even after. On
the other hand, $^{56}$Ni should be generated within short timescale
after the onset of the explosion, i.e. $\mathcal{O}(0.1)$ s, because
in order to synthesize $^{56}$Ni high temperature ($>5\times 10^9$ K)
is necessary and temperature decreases rather fast as the shock
propagates. Therefore, the amount of $^{56}$Ni is better indicator for
the explosion condition.

In this paper, we investigated the amount of $^{56}$Ni synthesized by
explosive nucleosynthesis in supernova ejecta by means of numerical
simulations and an analytic model. For numerical simulations, we
employ Lagrangian hydrodynamics code in which neutrino heating and
cooling terms are taken into account by light-bulb
approximation. Initial conditions are taken from \cite{woos07}, which
have 12, 15, 20, and 25 $M_\odot$ in zero age main sequence. We
additionally developed the analytic model, which gives a reasonable
estimate of the amount of $^{56}$Ni. We found that to produce enough
amount of $^{56}$Ni (0.07 $M_\odot$), we need $\mathcal{O}(1)$ Bethe
s$^{-1}$ of growth rate of the explosion energy, which is much larger
than canonical exploding simulations, typically $\mathcal{O}(0.1)$
Bethe s$^{-1}$.

It should be noted that a recent model fitting study suggested that
the distribution of $M$($^{56}$Ni) in normal type-II supernovae is
rather broad, i.e. from 0.005 to 0.28 $M_\odot$ \citep{muel17}. Our
model implies that these diversity can be mainly produced by different
progenitor masses, i.e. lighter progenitor models would produce less
$^{56}$Ni than more massive progenitors.  However, it should be also
noted that estimates of local supernovae are concentrating around 0.07
$M_\odot$ \citep[e.g.,][]{arne89}. With precise measurements of
$M$($^{56}$Ni) and the ejecta mass (related to progenitor mass), it is
able to give stringent constraint on the explosion mechanism of
core-collapse supernovae.  The current study also implies that in
order to produce enough amount of $^{56}$Ni, progenitor models which
have a large value of compactness parameter are preferred. This is
reasonable because a progenitor model, which has a small compactness
parameter, is extended and temperature of important mass coordinate
($\sim 0.1M_\odot$ above shock launching point) cannot be high enough
to synthesize $^{56}$Ni. This trend is opposite to the explodability,
which prefers a small value of compactness to produce the successful
explosion. These two observations may indicate that there is a limited
parameter space of progenitors, which can explain both the
explodability and $^{56}$Ni production simultaneously.

\section*{Acknowledgements}

This study was supported in part by the Grant-in-Aid for Scientific
Research (Nos. 26800100, 15H02075, 15H05440, 16H00869, 16H02158,
16H02168, 16K17665, and 17H02864). YS was supported by MEXT as
``Priority Issue on Post-K computer'' (Elucidation of the Fundamental
Laws and Evolution of the Universe) and JICFuS. TN and KM were
supported by the World Premier International Research Center
Initiative (WPI Initiative), MEXT, Japan. Discussions during the YITP
workshop YITP-T-16-05 on ``Transient Universe in the Big Survey Era:
Understanding the Nature of Astrophysical Explosive Phenomena'' were
useful to complete this work.

\end{document}